\documentclass[article]{aastex631}

\shorttitle{Gravitational lensing of Continuous GWs}
\shortauthors{}

\begin{document}
\title{Gravitational  lensing of Continuous Gravitational Waves}

\author[0000-0003-1308-7304]{Marek Biesiada}
\email{marek.biesiada@ncbj.gov.pl}
\affiliation{National Centre for Nuclear Research \\
Pasteura 7, 02-093 Warsaw, Poland \\
}
\affiliation{Department of Astronomy, Beijing Normal University,\\ Beijing 100875, People's Republic of China\\}
\author[0000-0002-2653-7282]{Sreekanth Harikumar}
\email{ sreekanth.harikumar@ncbj.gov.pl}
\affiliation{National Centre for Nuclear Research \\
Pasteura 7, 02-093 Warsaw, Poland \\
}



\begin{abstract}

Continuous gravitational waves are analogous to monochromatic light and therefore could be used to detect wave effects like interference or diffraction. This would be possible with strongly lensed gravitational waves. This article reviews and summaries the theory of gravitational lensing in the context of gravitational waves in two different regimes: geometric optics and  wave optics, for two widely used lens models such as point mass lens and Singular Isothermal Sphere (SIS). Observable effects  due to wave nature of gravitational waves are discussed. As a consequence of interference GWs produce beat patterns which might be observable with the next generation detectors like ground based Einstein Telescope, Cosmic Explorer or space-borne LISA, DECIGO. This will provide us a opportunity to estimate the properties of lensing system and other cosmological parameters with  alternative techniques. Diffractive microlensing could become a valuable method to search for intermediate mass black holes formed in the centers of globular clusters. We also point to an interesting idea of detecting the Poisson-Arago spot proposed in the literature. 

\end{abstract}



\section{Introduction} \label{sec:intro}

With the first detection of gravitational waves in 2015 \cite{PhysRevLett.116.061102} from coalescing compact binary system  we have entered the long expected era of GW astronomy. The new window on the Universe has been opened. First GW detections brought the confirmation of existence of binary black hole (BBH) systems in Nature. A half of century ago, primary candidates for chirping signals were binary neutron stars (BNS) due to sober expectations based on Hulse-Taylor like BNS systems discovered so far. Indeed in 2017  first such coalescence has been registered \cite{PhysRevLett.119.161101} and accompanying electromagnetic signals spanning from gamma-rays, through the optical to the radio waves has been registered allowing the identification of host galaxy and making a plethora of various tests possible. Further observing runs of LIGO-Virgo-KAGRA network enriched considerably the statistics of registered events. In the future there will be qualitative boost when the 3rd generation of ground-based detectors like the Einstein Telescope(ET) \cite{2010} or Cosmic Explorer(CE) \cite{PhysRevD.91.082001} as well as space-borne detectors (LISA, DECIGO, TianQin) become operative. First, the sensitivity of ground-based detectors will be increased by an order of magnitude over the existing ones, allowing to probe three orders of magnitude larger volume of the Universe. Second, the satellite detectors will probe much lower frequencies of GW (inaccessible from the ground due to seismic noise) enabling to observe adiabatic inspiralling signals from binary systems much earlier than the coalescence phase probed by ground based detectors. This means that besides the already registered chirp signals, we would gain access to almost monochromatic continuous GW signals. 

In this paper we discuss some new opportunities, which will be available when continuous GW signals are registered. In the case of light, historically there has been a dispute about its nature: corpuscular vs. wave. Wave nature of light has been established with the interference and diffraction patterns, which are best visible with monochromatic light. By analogy, interference and diffraction patterns should be revealed in GWs as well. But what sort of experiment could be done in this context? We point to gravitationally lensed GW signals as a promising physical setting to reveal wave effects of GWs.  

Bending of light by the Sun was the first classic test of General Relativity. The essence of light bending by massive bodies lies in the fact that paths of photons (light rays) are null geodesics in spacetime, which is curved by the presence of mass. Since the Eddington's expedition in 1919, this phenomenon now evolved to a mature discipline of extragalactic astronomy called gravitational lensing.  Currently it is routinely used to study the structure of galaxies (lenses) and cosmology. Gravitational waves, in the geometric optics approximation, follow the null geodesics as well.  Hence one can expect to see gravitationally lensed GW signals. Such signals coming from unresolved images would interfere producing characteristic patterns. In the next section we will review strong lensing of GWs. Section 3 contains discussion of some observable effects particularly pronounced when the source emits continuous GWs. We summarize in Section 4.
\section{Gravitational Lensing of Gravitational Waves}

As one of the successful predictions of General Relativity, strong
gravitational lensing by galaxies has become one of the most
important tools in studying cosmology \cite{Cao15}, galaxy structure and evolution \cite{Cao2016}. With the dawn of gravitational wave
astronomy, the robust prediction suggested that a considerable
number of GW signals from inspiralling neutron stars would be
gravitationally lensed, focusing on the third generation of
ground-based GW detectors \cite{Ding15} and second generation of
space-based GW detectors \cite{Piorkowska2021}, respectively. 

Strong gravitational lensing occurs whenever the source, the lens (a galaxy, a black hole, a star) and the observer are almost aligned. In the particular case of perfect alignment the source will not be obscured by the lens but instead visible as a glowing ring -- called the Einstein ring. In the case of axisymmetric lenses, the angular radius of this ring i.e. the Einstein radius is:
\begin{equation}
{\theta _E} = \sqrt {\frac{{4G{M_l}}}{{{c^2}}}\frac{{{D_{ls}}}}{{{D_l}{D_s}}}} ,
\end{equation}
where $M_l$ is the mass of the lens, ${D_l}$ is the angular diameter distance to the lens at redshift ${z_l}$, ${{D_s}}$ is the angular diameter distance to the source at redshift ${z_s}$, ${D_{ls}}$ is the angular diameter distance between the lens and the source. 
Misalignment in this optical system by the angle $\beta$ (angle between directions to the source and to the lens) leads to the appearance of multiple images. Usually two (spherically symmetric mass distribution) or four images are formed at angular separations of $\theta_i$ from the center of the lens. The typical separation between different images is set by the Einstein radius and their locations is described by the lens equation: 
\begin{equation} \label{lens eq}
    \beta = \theta - \alpha(\theta)
\end{equation}
where: $\alpha(\theta)$ is the deflection angle determined by the projected mass distribution of the lens (for detailed derivations see \cite{1992Schneider}).   
The total time delay introduced by gravitational lensing at the angular position $\theta$ from the lens is:
\begin{equation} \label{time delay functional}
    \Delta t = \frac{1+z_l}{c} \frac{D_lD_s}{D_{ls}} \left[ \frac{(\theta - \beta)^2}{2} - \phi(\theta) + \phi_m(\beta) \right]
 \end{equation}
  where $\phi(\theta)$ is the lens potential determining the deflection angle 
 $\alpha(\theta) = \nabla_{\theta} \phi(\theta)$. Term  $\phi_m(\beta)$ corresponds to the arrival time in a non-lensed case and in practice it is a constant adjusted to ensure the extreme value of time delay functional. Total time delay, more precisely the time delay functional measures the delay between lensed signal arrival from the image at position $\theta$ with respect to (unmeasurable) arrival time from the source if the lens was absent. Fermat's principle implies that images correspond to stationary points of time delay functional, and one can see that condition $\nabla_{\theta} \Delta t = 0$ is equivalent to the lens equation (\ref{lens eq}). 
The measurable quantity is the time delay difference $\Delta t_{ij}$ between images at $\theta_i$ and $\theta_j$. Another measurable quantity is the ratio of image magnifications. The (signed) magnification is the inverse Jacobian of the lens equation: $\mu = \left(det\left(\frac{\partial \beta}{\partial \theta}\right) \right)^{-1}$. The sign corresponds to the parity of the image, so physically relevant are $|\mu|$ and since usually the intrinsic luminosity is unknown, only magnification ratios are relevant. The exception could be the lensed standard candles. In the optical, image locations are another sort of observables, but for the purpose of our discussion they are irrelevant. 

For the purpose of this work, we refer to two simplest, commonly used, analytic lens models: point mass model and singular isothermal sphere (SIS) model. Their choice is dictated by two realistic scenarios: lensing by an isolated mass like a massive (with $M_l<10^{7} M_{\odot}$) black hole (BH) or an intervening galaxy ($M_l \sim 10^{9} - 10^{11} M_{\odot}$).

In the case of axisymmetric point lenses, the typical separation between different images is set by the Einstein radius
\begin{equation}
{\theta _E} = \sqrt {\frac{{4G{M_l}}}{{{c^2}}}\frac{{{D_{ls}}}}{{{D_l}{D_s}}}} ,
\end{equation}
where $M_l$ is the mass of the lens, ${D_l}$ is the angular diameter distance to the lens at redshift ${z_l}$, ${{D_s}}$ is the angular diameter distance to the source at redshift ${z_s}$, ${D_{ls}}$ is the angular diameter distance between the lens and the source. 
The dimensionless lens equation \citep{1992Schneider} and its corresponding solutions are
\begin{equation}
y = x - \frac{1}{x}\rightarrow \begin{array}{*{20}{c}}
{\left\{ {\begin{array}{*{20}{c}}
{{x_ + } = \frac{{y + \sqrt {{y^2} + 4} }}{2}}\\
{{x_ - } = \frac{{y - \sqrt {{y^2} + 4} }}{2}}
\end{array}} \right.}
\end{array}
\end{equation}
where $y = \beta /{\theta _E}$, $\beta$ denoting the angle between directions to the lens and to the source,  $x = \theta /{\theta _E}$, where $\theta $ is the angular position of the image actually seen by the observer. 

In the case of point mass lenses, two images could be produced at angular positions ${\theta _ + } = {x_ + }{\theta _E}$ and ${\theta _ - } = {x_ - }{\theta _E}$, respectively. The inverse of the determinant of the Jacobi matrix $\frac{\partial \beta}{\partial \theta}$ defines their magnifications,
\begin{equation}
{\mu _ \pm } = \frac{1}{1 - {\left(\frac{1}{{{x_ \pm }}}\right)^4}}.
\end{equation}
Time delay between images produced by a point mass lens is \cite{1992Schneider}:
\begin{equation}\label{pm_td}
\Delta t = \frac{{4G{M_l}(1 + {z_l})}}{{{c^3}}} \left( \frac{{y\sqrt {{y^2} + 4} }}{2} + \ln \frac{{\sqrt {{y^2} + 4}  + y}}{{\sqrt {{y^2} + 4}  - y}} \right).{\rm{ }}
\end{equation}
Point mass lenses (also called Schwarzschild lenses) are representative to lensing by stars (microlensing) and BHs with stellar or intermediate mass BHs.

It is well established that early-type galaxies act as lenses in the majority of strongly gravitational lens systems detected. Even though their formation and evolution are still not fully understood in details, a singular isothermal sphere model (SIS) can reasonably characterize the mass distribution of massive elliptical galaxies within the effective radius
\citep{Treu04,Treu06,Cao2016,Liu20}. In the case of axisymmetric SIS model \citep{nara96,bern99}, its three-dimensional density profile could be described as $\rho (r) = {{\sigma _v^2}}/{{2\pi G{r^2}}}$, where $r$ is the distance from the sphere center and ${\sigma _v}$ is the velocity dispersion of the lens. 
In this case, the dimensionless lens equation \citep{1992Schneider} could be written as
\begin{equation}
y = x - \frac{x}{{|x|}},
\end{equation}
where $x = \theta /{\theta _E}$, $y = \beta /{\theta _E}$ as above.  
There are two cases for the solution of the lens equation: when $y < 1$, the solutions are
\begin{equation}
x _ {\pm}  = y \pm 1.
\end{equation}
The corresponding angular positions of the images are  ${\theta _ \pm } = \beta  \pm {\theta _E}$, where the Einstein radius ${\theta _E}$ could be written as
\begin{equation}
{\theta _E} = \sqrt {\frac{{4GM(\theta _E)}}{{{c^2}}}\frac{{{D_{ls}}}}{{{D_l}{D_s}}}} =  4 \pi \frac{\sigma_{v}^2}{c^2}
\frac{D_{ls}}{D_s},
\end{equation}
where $M(\theta _E)$ is the mass within the Einstein radius. In this case, the magnifications of the images are
\begin{equation}
\mu _ {\pm}  = 1 \pm \frac{1}{y}.
\end{equation}
On the other hand, when $y > 1$, the lens equation only has one solution: $x=y+1$. Its corresponding magnification is $\mu  = {{|x|}}/({{|x| - 1}})$.

For SIS model, the time delay between different images reads:
\begin{equation} \label{sis_td}
\Delta t = \frac{32 \pi^2}{c} \left(\frac{{{\sigma _v}}}{c}\right)^4  \frac{{{D_l}{D_{ls}}}}{{{D_s}}}(1 + {z_l})y.
\end{equation}

\subsection{Wave optics regime}

The above summarized results have been obtained using the geometric optics approximation, which is excellent is most astrophysically relevant cases of strong gravitational lensing of electromagnetic waves. Indeed this approximation is valid as long as typical length scale of the system is much larger than the wavelength and timescale of typical variations of the system are much larger than the period of the wave. In the case of GWs it becomes different:  frequency range probed by ground-based detectors comprise $10\;Hz<f<10\;kHz$ (to be lowered down to $1\;Hz$ in the future third generation) and in the space-borne detectors it will be $0.1\;mHz<f<100\;mHz$ in LISA and $1\;mHz<f<100\;Hz$ in DECIGO. This corresponds to the GW wavelenghts of $10^4\;m< \lambda <10^7\;m$ in ground-based and  $10^6\;m< \lambda <10^{12}\;m$ in space-borne detectors. The first papers acknowledging the importance of wave regime for GW lensing were \cite{Nakamura98, Takahashi03}.

In the wave optics regime, we consider  the gravitational wave from a distant source as  propagating in the background spacetime $g_{\mu\nu}^{(L)}$ of the lens object characterized by the gravitational potential $U(\textbf{r}) << 1$. The total metric, including the  perturbation (due to GW signal)  is given by $g_{\mu\nu} =  g_{\mu\nu}^{(L)}+ h_{\mu\nu}$, where $h_{\mu \nu} = \psi e_{\mu \nu}$. Since the polarization tensor $e_{\mu \nu}$ is parallel transported along the null geodesic, we may neglect its change (being of order $\sim U(\textbf{r}) << 1$) and study the propagation equation of the scalar wave: $\partial_{\mu} \left( \sqrt{-g^{(L)}}g^{(L)\mu \nu} \partial_{\nu} \psi \right) = 0 $. For the monochromatic scalar wave $\psi(\textbf{r},t) = \tilde{\psi(\textbf{r})} e^{-2 \pi i f t}$, in the frequency domain and remembering the weak potential assumption, this equation can be cast to the form of Helmholtz equation:
\begin{equation}
 \left( \nabla^2 + 4 \pi^2 f^2 \right) \tilde{\psi} = 16 \pi^2 f^2 U \tilde{\psi}
\end{equation}

The solution to the Helmholtz equation can be given in terms of Kirchoff integral \cite{1992Schneider}, where it is convenient to introduce the dimensionless amplification factor: 
\begin{equation}
    F(f) =  \frac{\tilde{\psi}^{L}}{\tilde{\psi}}
\end{equation} \label{amplification factor}
which is the ratio of wave amplitudes with and without lensing.Then the Kirchoff integral allows us to calculate the amplification factor at the observer \cite{1992Schneider} as:
\begin{equation} \label{Kirchoff}
    F(f,\beta) =  \frac{1+z_l}{c} \frac{D_{s}}{D_l D_{ls}} \frac{f}{i} \int d^{2}\theta \exp\left[ 2\pi i f \Delta t(\theta, \beta) \right]
\end{equation}
where $\Delta t$ is lensing time delay given by equation (\ref{time delay functional}) and cosmological nature of distances in the optical system has been taken into account.  Calculation of the integral (\ref{Kirchoff}) can be simplified by switching from angles to dimensionless variables $x$ and $y$ (see above after Eq.(5)) and introducing new parameter $w = \frac{8 \pi G}{c^3} M_l(\theta_E) (1+z_l) f$:
\begin{equation} \label{Fw}
    F(w,y) = \frac{w}{2 \pi i} \int d^2y \left[ \exp[ i w T(x,y) \right]
\end{equation}
where: $T(x,y) = \frac{(x-y)^2}{2} - \frac{\phi(x)}{\theta_E^2}$. In general amplification factor (\ref{Fw}) should be evaluated numerically, but in two particular cases it can be integrated analytically.  
In the point mass lens model the dimensionless lensing potential is $ \phi(\textbf{x}) = ln x$ and the analytical solution to the diffraction integral is  given by 
\begin{equation}
    F(f) = \exp \Big [  \frac{\pi w}{4} + i\frac{w}{2}\Big( ln\Big( \frac{w}{2} \Big) -2 \phi_{m}(y)\Big) \Big] \Gamma \left( 1-\frac{i}{2}w \right)\; {}_1F_{1}\left(  \frac{i}{2}w,1;w y^{2} \right)
\end{equation}
where $\phi_{m}(y) =  0.5(x_{m}- y)^{2}/2 - ln x_{m}$ with $x_{m} =  (y+ \sqrt{y^{2}+4})/2$  and ${}_1F_{1}$ is the confluent hypergeometric function.
In the case of SIS model, the Kirchoff integral could be cast into analytically manageable form \cite{Takahashi03}:
\begin{equation}
    F(f) =  -iw \; e^{iwy^{2}/2} \int_{0}^{\infty} dx \; x J_{0}(w xy) \exp \left[ i w \left(  \frac{1}{2}x^{2}-x + \phi_{m}(y) \right)  \right]
\end{equation}
where $\phi_{m}(y) =  y + 1/2$ and the $J_{0}$ is the Bessel function of order zero. 
Geometric optics corresponds formally to $f \to \infty$, hence $w >> 1$. In such case Kirchoff integral can be calculated using stationary phase approximation and dominant contributions will come from image positions. Hence the amplification of the wave intensity is: 
\begin{equation}\label{ampfac1}
    |F(f)|^2 = |\mu_{+}|^2 + |\mu_{-}|^2 + 2 |\mu_{+} \mu_{-}| \sin{(2\pi  f \Delta t_{d})}
\end{equation}
where first two terms correspond to magnifications in geometric optics, the last one represents the interference between images. 

\section{Effects of gravitational wave lensing of continuous waves}

Since the lensed images would not be resolved, they are expected to come from the same location on the sky separated in time by the lensing time delay $\Delta t_{ij}$. In the case of compact binary systems coalescences this has been expected to be revealed as just two signals with similar temporal patterns and amplitudes affected by magnification. This picture is basically true for merging systems staying within the detectors' sensitivity band for a fraction of a second. However, if such a system could be monitored earlier i.e. in adiabatic inspiral phase (almost monochromatic signal), then the brighter image (arriving earlier) could imprint its interference patterns on the strain of the fainter image. In this section we will discuss some of these effects stemming from the wave nature of GWs, which could be observed. We review here some fresh ideas that already appeared in the literature (and to which one of us contributed).  
\subsection{Beat Patterns}

In an interesting paper \cite{Hou2020} it has been discussed that if the lensed images from the coalescing binary system could be detected simultaneously by the detector, then beat patterns would be revealed prior to the merger signal. The merger signal would arrive first from the brighter (more magnified) image, then after lensing time delay the merger signal from the fainter one would be registered. If they stayed for a couple of seconds in the detector sensitivity band, then one would see beat patterns. First of all, this condition is very restrictive for the strong lensing system configuration: time delay should be of order of minutes at most. This means that probability of detecting such an event from the ground is much lower than probability of detecting a strong lensed GW in general. However, the payoff of such detection should be considerable, because as the authors of \cite{Hou2020} argue, one would be able to measure the redshifted lens mass purely from the GW signal, and measure time delay distance without reconstructing the Fermat potential. This is interesting enough, so that it is worth to recall the basic points here. 

Suppose that we have a signal in time domain $h(t)$, which is a superposition of two differently magnified signals from unresolved two images: 
\begin{equation}
h = \mu_{+}[A^{+} \cos(\omega_{1}t + \phi_{1}) + A^{\times} \sin(\omega_{1}t + \phi_{1})  ] + \mu_{-}[ A^{+} \cos(\omega_{2}t + \phi_{2}) + A^{\times} \sin(\omega_{2}t + \phi_{2})]
\end{equation}
where $\mu_{+}$ and $\mu_{-}$ are their respective magnification, $A^{+}$, $A^{\times}$ are the amplitudes and $\omega_{1}$,  $\omega_{2}$ , $\phi_{1}$, $\phi_{2}$ are angular frequencies and initial phases respectively.
It can be rewritten as:
\begin{eqnarray} \label{beat}
    h = \mu_{sum} \left[A^{+} \cos(\omega_{f}t + \phi_{f}) \cos(\omega_{b}t + \phi_{b}) + A^{\times} \sin(\omega_{f}t + \phi_{f}) \cos(\omega_{b}t + \phi_{b})  \right] +  \nonumber \\ 
    + \mu_{diff} \left[ - A^{+} \sin(\omega_{f}t + \phi_{f}) \sin(\omega_{b}t + \phi_{b}) + A^{\times} \cos(\omega_{f}t + \phi_{f})  \sin(\omega_{b}t + \phi_{b} )   \right] \nonumber \\
    &&
\end{eqnarray}
with $\mu_{sum} = \mu_{+}+\mu_{-},$ $\mu_{diff} = \mu_{+} - \mu_{-},$ the frequencies $\omega_{b}=\frac{\omega_1-\omega_2}{2}$ and $\omega_{f}=\frac{\omega_1+\omega_2}{2}$ represent beat and average frequency, respectively and phases $\phi_b, \; \phi_f$ are defined in a similar manner. In the early inspiral phase (i.e. well before coalescence) when the frequency evolves according to $\frac{d\omega}{dt} = \frac{192}{5} {\cal M}^{5/3} \left(\frac{\omega}{2}\right)^{11/3}$ \cite{jaranowski-krolak, Maggiore2009} and $\omega_b<<\omega_f \approx \omega_1 \approx \omega_2$ one has:
\begin{equation}
    \omega_{b} =  \frac{96}{5} \Big(  \frac{\omega_{f}}{2} \Big)^{11/3} \mathcal{M}\Delta t
\end{equation}
where $\mathcal{M}$ is the chirp mass of the binary system and $\Delta t$ is the lensing time delay (\ref{time delay functional}) given by formula (\ref{pm_td}) in case of point mass lens. Fig. 1 illustrates the beat pattern originating in the lensed coalescing system. One can see the evolution of beat pattern toward the coalescence. During the last stages of inspiral $\omega_{f}$ increases and the beat pattern disappears. In the case of lensed continuous GW signals beat pattern would be rigid. Now, if one use beat pattern defined by equation (\ref{beat}) as a template in matched filtering method, then the ratio $\frac{\mu_{sum}}{\mu_{diff}}$ can be extracted from the data and easily converted to the magnification ratio $r=\left|\frac{\mu_{+}}{\mu_{-}}\right|$. The lensing time delay $\Delta t$ is directly measurable and in the case of registering two coalescence signals the accuracy of such measurement would be unprecedented. In the case of almost monochromatic signals from the inspiral phase long before the merger (accessible in low frequency bands of DECIGO and LISA) time delay $\Delta t$ could still be extracted from the matched filtering using beat pattern template.  
The knowledge of the magnification ratio $r$ will allow to make interesting inferences. In the case of a point mass lens, after expressing time delay (\ref{pm_td}) in terms of $r$, one can see that redshifted mass of the lens can be calculated as:
\begin{equation} \label{pm mass}
    M_l (1+z_l) = \frac{c^3}{4 G} \Delta t \left[ \frac{r-1}{\sqrt{r}} + \ln r\right]^{-1}
\end{equation}
On the other, hand in the case of SIS lens the so called time delay distance $D_{\Delta t}$ can be obtained:
\begin{equation}
    D_{\Delta t} \equiv \frac{{{D_l}{D_{ls}}}}{{{D_s}}} = \frac{c^5 \Delta t}{32 \pi^2 \sigma_v^4 (1+z_l)} \frac{r-1}{r+1}
\end{equation}
provided that lens is identified and its redshift is known. This is also an important result, demonstrating the breadth of information which can be extracted from GW signal alone. Needless to say that time delay distance measurements have been used to determine the Hubble constant in a way independent of the cosmic ladder or CMB techniques. It has been shown in \cite{NatureComm} that lensed GW signals from coalescing BNS systems would allow a percent accuracy in $H_0$ measurements. Moreover, time delay distance can be used to test the cosmological model beyond the $\Lambda$CDM. 
\begin{figure}[h]

\includegraphics[width=8cm, height=6cm]{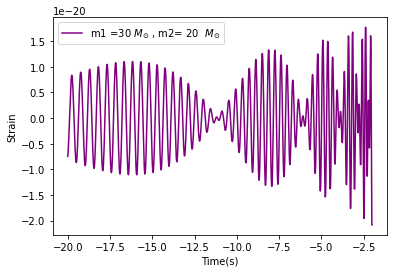}
\centering
\caption{A schematic diagram showing the beat pattern for a binary inspiral signal. Masses of the components are $m_{1} =  30 M_{\odot}$ and $m_{2} =  20 M_{\odot}$, source is located at $z_s = 2$, lens at $z_l = 1$ is assumed to be a point mass of  $M_l = 10^6\;M_{\odot}.$ Time delay has been chosen as $\Delta t = 2.1\;s$ thus fixing the $\beta$ parameter respectively. }
\label{fig:image2}
\end{figure}

\subsection{Diffractive microlensing}

Besides the strong lensing discussed in previous section, which in case of electromagnetic window usually means multiple images of a quasar (or galaxy) lensed by a galaxy, one can consider microlensing. It is again a strong lensing case (usually starlight lensed by a star) where the images cannot be resolved, but the relative motion of the elements of the optical system make temporal changes of  total magnification of the source detectable. It was an ingenious idea of Paczy{\'n}ski \cite{pacz1986} that two individually unobservable effects (image separation and relative proper motion) can be combined to give observable effect. One can expect, that the similar phenomenon could be detected in GW domain. In the paper \cite{liao2019} such setting has been proposed and discussed. Besides \cite{liao2019} fringes due to GW lensing have also been studied in  \cite{Lai2018,Jung2019,Christian2018}. 
One can think of gravitational lensing of GWs (in or close to wave regime) as an analog of one-slit or  two-slit experiment with light. When the light is monochromatic (a laser beam) a characteristic fringe pattern is visible on the screen. In astrophysical scales the ``screen'' has astronomical scales and is inaccessible directly. This corresponds to fixed $y$ parameter. However, in the microlensing case $y$ changes in time (due to relative motions) and one can explore the ``screen'' by watching the manifestation of fringes.
The peculiar motions of the lensing system mentioned above include the motion of the observer $\textbf{v}_{obs}$, the source $\textbf{v}_{s}$ and the lens $ \textbf{v}_{l} $- all measured in the CMB frame combined to obtain the effective velocity $\textbf{v}_{eff}$ as given by \cite{Kayser1986}
\begin{equation}
    \textbf{v}_{eff} =  \textbf{v}_{s} - \frac{1+z_{s}}{1+z_{l}}\frac{D_{s}}{D_{l}}\textbf{v}_{l} +\frac{1+z_{s}}{1+z_{l}}\frac{D_{ls}}{D_{l}}\textbf{v}_{obs}
\end{equation}
Since in lensing we are referring to angles on the sky, the effective velocity should be converted to angular proper motion.  

For the comparison between the size of ``barrier'' (lens) and the wavelength of GW one uses of the parameter $w=\frac{8 \pi G}{c^3} M_l(\theta_E) (1+z_l) f$ introduced in section 2. Recall that if $ w >> 1 $ the amplification factor in (\ref{Kirchoff}) simplifies to (\ref{ampfac1}). The last term in (\ref{ampfac1}) is the interference term.
If the amplitude of the diffraction fringes are to be detectable the parameter $w$ should be large enough (but not too big). It has been proposed in \cite{liao2019} that $w>1$ is the reasonable condition for the fringe pattern to be detectable. To see a complete diffraction or interference pattern the observation period $T_{obs}$  should  be more than the timescale over which the fringe pattern are affected due to peculiar motion of the system, we denote the fringe timescale as $t_{f}$. This timescale of transit of fringes can be denoted by $t_{f} = \frac{t_{E}}{w}$ where $t_{E}$ is the Einstein radius crossing time. For galactic sources and lens it is given by 
\begin{equation}
    t_{E} \approx 34.7 \text{days}\sqrt{4\frac{D_{l}}{D_{s}}\Big(  1-\frac{D_{l}}{D_{s}} \Big)}\Big(\frac{D_{s}}{8 \text{kpc} }\Big)^{1/2} \times \Big( \frac{M}{M_{\odot}} \Big)^{1/2} \Big(  \frac{v_{eff}}{200 \text{km}s^{-1}} \Big)^{-1}
\end{equation}
Therefore the criterion for detecting a fringe pattern can be summarized in three conditions \cite{liao2019}: \begin{enumerate}
\item w $>$1, to have enough amplification
    \item y$<$3 to detect fringes before they are damped
    \item $\Delta t > t_{f}$ to see a fringe pattern
\end{enumerate}
Let us observe, that the second condition means that the probability of microlensing is almost an order of magnitude larger than in the optical. It is because the cross-section for lensing is proportional to the Einstein ring area $\pi y^2$ and in the optical $y=1$ is the threshold for microlensing.  
Estimates of \cite{nar90,arnett} suggest that galactic bulge contains about $10^{9}$ NSs and globular cluster contain $10^{3}$ NSs which could be the target population for the current and next generation gravitational detectors. In order to find the probability of observing fringe pattern from such a population of monochromatic sources one need to calculate optical depth which gives the probability that a given source falls
into the Einstein radius of any lens located along the line of
sight $ \tau =  \int_{0}^{D_{s}}n(D_{l})\pi R_{E}^{2}d D_{l}.$ 
If we consider a source population in the Milky Way bulge lensed by a constant density of stellar mass lenses in the galactic disk moving with a velocity $v_{rot} =  200 \text{kms}^{-1}$, the lensing probability is given by \cite{pacz1986,pacz1991}.
\begin{equation}
    \tau =  \frac{v_{rot}^{2}}{2c^{2}} =  f_{l}\times 10^{-6}
\end{equation}
where $f_{l}$ is the fraction of lens larger than one solar mass, which is required to fulfill the above mentioned conditions for diffractive microlensing. 
Currently, the issue of intermediate mass BHs and their existence has been debated. They can be formed in the very cores of globular clusters. Therefore, it would be interesting to focus on globular clusters as hosts of gravitational lenses for lensing continuous GWs. One such candidate is already known. M22 is a rich globular cluster with projected location pointing the Galactic bulge -- a perfect site of sources (spinning neutron stars or binary systems). A typical velocity dispersion in a globular cluster is $\sigma_{v}\sim 10 $ km$s^{-1}$, hence the typical transverse velocity of the lens can be very well approximated by that of a cluster as a whole $ \sim 200\; \text{km s}^{-1}$. Again, \cite{Paczynski1994} has shown that the optical depth for microlensing in such case is: 
\begin{equation} \label{M22}
    \tau = f_l \frac{\sigma^2}{c^2} \frac{2 \pi}{\varphi} \frac{D_{ls}}{D_s} \approx 2.4 \;\; 10^{-4} \; f_l \left( \frac{\sigma_v}{10\;km s^{-1}} \right)^2 \left( \frac{1'}{\varphi} \right) \frac{D_{ls}}{D_s}  
\end{equation}
where $\varphi$ denotes the angular distance from the center of the cluster. Knowledge of the distance and transversal velocity of lenses belonging to the cluster may result with accurate estimates of their masses, breaking the degeneracies inherent to the microlensing technique. The microlensing of the bulge star at $D_s = 8.2 \; kpc$ by a low-mass object in the globular cluster M22 located at $D_l = 2.6\; kpc$ has been reported in e.g. \cite{Pietrukiewicz}. Let us remark that the inclusion of $\varphi$ term has been motivated by the fact that in the optical only the outskirts of the cluster can produce microlensing of the bulge stars. Regions close to the center would obscure the light of bulge stars. In the GW domain this is completely different - GWs can probe central parts of the cluster easily. This will significantly enhance the optical depth. Moreover, the intermediate mass black holes  (IMBHs) $M \sim 10^2 - 10^4\;M_{\odot}$ are expected to reside in the centers of globular clusters. Hence, diffractive microlensing could open a new chapter of searching for IMBHs. One of the candidates of continuous GW signals is a (slightly deformed) spinning neutron star. Their detection is fairly difficult, since the signal is strongly modulated by the Earth’s rotation and orbital motion. Moreover, this modulation is different for every sky position. Diffraction and interference fringes caused by intervening mass acting as a lens and moving
with respect to the source also produce modulation. Luckily, the timescales of these effects are significantly different from timescales involved with the motion of the detector. In the setting discussed by us here, the directed search with the know position on the sky (in case of globular cluster lensing) can be applied. The GW signal coming from a rotating
NS is so weak that, in order to detect it in the detector’s noise, one has to analyze months-long segments of data. Implementation of the F-statistic \cite{jaranowski-krolak} consists of coherent searches over two-day periods, and is followed by a search for coincidences among the candidates from the two-day segments. The timescales for fringe modulation are larger, and one can expect that
respective amplitude modulation could be detectable with current techniques. This approach should be extended to other sources of continuous GWs like binary systems observable in low-frequency bands to be probed by DECIGO and LISA. 
\subsection{Poisson-Arago spot}
One of the observable signature of wave nature of light besides interference is diffraction. The French physicist Fresnel presented his work on diffraction in a competition sponsored by the French Academy of Sciences in 1818. It was a period when Newtons idea of corpuscular theory of light was prevalent, one of the judges in the  competition Poisson in an attempt to disprove the wave nature of light has pointed out that according to Fresnel's theory a parallel beam of light falling on a spherical obstacle would produce a bright spot at the centre as if the obstacle is not present. This was considered to be an absurd result, later another French physicist Arago  successfully conducted an experiment and the Poisson spot was seen. This phenomenon later came to be known as the Poisson's spot or the Spot of Arago established the wave nature of light in 19th century. Therefore, the similar phenomenon should be expected for GWs and strong lensing optical configuration appears to be a perfect candidate. The reason for the appearence of the spot is that wavefronts originating at the boundary of circular obstacle interfere constructively. The same is expected from the Einstein ring of lensed GW.
To some extent one can see analogous behavior in wave GW regime, when the amplification calculated from the Kirchoff integral (\ref{Fw}) has maximal value, when $\beta=0$ \cite{Oguri}
\begin{equation}
    \left| F(w,\beta=0) \right|^2 = \frac{\pi w}{1 - e^{-\pi w}}.
\end{equation}
An interesting and rigorous study of the Poisson-Arago spot produced when GWs are passing in the background of a Schwarzschild BH has been presented in \cite{Hongsheng:2018ibg}. This strong field scenario can be understood as an on-axis scattering problem using the Regge-Wheeler equation\cite{Regge-wheeler, scattering19} which describes the axial perturbations of the Schwarzschild metric as linear approximation\cite{Plamen06} and the solutions are given by confluent Heun equation. The amplitude of the gravitational wave passing in the background of this opaque disc is given by 
\begin{equation}\label{radialeqn}
    \Phi_{l} =  \frac{1}{r}\sum_{l=2}^{\infty}(2l+1)e^{-i\omega t}P_{l}(\cos\theta)R_{ls}(r)
\end{equation}
where $P_{l}(\cos\theta)$ denotes the Legendre polynomials and $R_{ls}$ is the $l-th$ partial wave of the GW with spin $s =2$ and angular momentum
$\hbar \sqrt{l(l+1)} $ measured with respect to the scattering center.
For a Schwarzschild BH the gravitational waves with impact parameters $3\sqrt{3}M$ falls into the black hole and will not reach infinity i.e) BH can shield gravitational waves.$R_{ls}/r$ in (\ref{radialeqn}) is an oscillatory function in the far zone at which the summation in $l$ in eqn(\ref{radialeqn}) becomes divergent. This issue can be solved in \cite{Hongsheng:2018ibg} by using the Fresnel half wave zone according to which effective contribution to diffraction comes from the $1/4-$ zone and the residues cancel each other. In the case of  small scattering angle and at large distance  $R_{ls}/r$ becomes 
\begin{equation}
   \Phi_{l} =  e^{-i\omega t}e^{\omega r (1-\frac{\theta^{2}}{4})+ \pi M \omega} \Gamma (1-2i\omega M)J_{0}\Big( 2\sqrt{\frac{M}{r}\omega r\theta}\Big)
\end{equation}
The fringe pattern is determined by the function $J_{0}\Big( 2\sqrt{\frac{M}{r}\omega r\theta}\Big)$. The time delay between successive bright or dark spots depends on the mass of the BH, wavelength of the gravitational wave and the relative position of the system. The most promising obstacle we expect to see Arago spot is the SMBH located at the centre of our Milky Way galaxy which acts like an opaque disc. For gravitational wave  of frequency 100 Hz the diffraction time delay between bright spot and the first dark fringe is 3.86 days and the second bright fringe is 6.2 days. These effects will be observable with the next generation space based detectors like LISA.

\section{Conclusions}
 
Gravitational lensing of GW signals has been discussed in many papers since the first detection \cite{PhysRevD.97.023012}. 
In particular, \cite{Smoot2019} put forward an idea that unexpectedly high masses of BBHs detected by LIGO are consequences of the signals being lensed. These claims have been subsequently refuted \cite{LIGO_lensed_GW}, but the interest in lensed GW remained. Some teams undertook efforts to monitor ongoing LIGO-Vrigo-KAGRA detections for the possibility of some signals being lensed by galaxy clusters \cite{Lensed_GW_hunters} with hope to repeat the Refsdal supernova story \cite{Refsdal}. 

In this paper we have discussed new opportunities of detecting wave effects in lensed GW signals like the beat patterns (interference) and diffractive microlensing. The former would allow independent measurement of the lens mass and the time delay distance purely from the GW signal. The latter would open a new opportunity to search for IMBHs inside globular clusters and more generally, study the central parts of globular clusters -- a technique completely inaccessible in the electromagnetic window.
The future 3rd generation of detectors like the Einstein Telescope or Cosmic Explorer is expected to detect lensed GW signals routinely at the rate of 50-100 events per year \cite{marek, Lilan2019}. The prospects are also bright for the planned space-borne detectors like DECIGO, which will probe an order of magnitude larger volume of the Universe that the ET. In \cite{Piorkowska2021} it has been estimated that DECIGO would routinely register about $50$ lensed GW signals per year. Remarkably, some of these sources will later enter the band of ground-based detectors. Inspiralling sources detectable by the DECIGO would exhibit slowly evolving frequency hence being almost monochromatic and allowing to detect patterns discussed in this paper. The expected rate of detecting beat patterns by DECIGO has been assessed in \cite{Decigo_beat}.



\begin{acknowledgments}

\end{acknowledgments}

%





\bibliography{sample631}{}
\bibliographystyle{aasjournal}



\end{document}